%% file: neuro-cs.tex
\documentclass[sigconf, nonacm]{acmart}
\usepackage{url}
\usepackage{color}
\usepackage{xspace}
\usepackage{caption}
\usepackage{subcaption}
\usepackage{adjustbox}
\usepackage{enumitem}
\usepackage{xspace}
\usepackage{etoolbox}
\usepackage{tcolorbox}
\usepackage{balance}
\usepackage{balance}

\newcommand{\eg}{\textit{e.g.}\xspace}
\newcommand{\etal}{\textit{et. al.}\xspace}
\newcommand{\ie}{\textit{i.e.}\xspace}
\newcommand{\etc}{\textit{etc.}\xspace}

\newcommand{\sr}{\textsf{SR}\xspace}
\newcommand{\nr}{\textsf{NR}\xspace}
\newcommand{\sentp}{\textsf{SP}\xspace}
\newcommand{\cp}{\textsf{CP}\xspace}
\newcommand{\scratch}{\textsf{ScratchJr}\xspace}
\newcommand{\codej}{\texttt{codeJ}\xspace}
\newcommand{\codee}{\texttt{codeE}\xspace}
\newcommand{\sentt}{\texttt{sent}\xspace}

\newcommand{\md}{MD}
\newcommand{\lrspace}{Language system\xspace}
\newcommand{\mdspace}{\md\xspace system\xspace}
\newcommand{\lrmd}{\md\xspace and the Language systems}
\newcommand{\lrmdspace}{\lrmd\xspace}

\newcommand{\codecontent}{\textit{code simulation}\xspace}
\newcommand{\codecontentcaps}{\textit{Code simulation}\xspace}
\newcommand{\codecomp}{\textit{code comprehension}\xspace}
\newcommand{\codecompi}{{code comprehension}\xspace}
\newcommand{\pu}{\textit{code reading}\xspace}
\newcommand{\putasks}{{coding tasks}\xspace}
\newcommand{\delbeta}{$\Delta\beta$}

\newcommand{\floyd}{Floyd \etal}
\newcommand{\siegmund}{Siegmund \etal}
\newcommand{\liu}{Liu \etal}
\newcommand{\subsec}[1]{\par\noindent\textbf{#1}}

\newcommand\blfootnote[1]{%
	\begingroup
	\renewcommand\thefootnote{}\footnote{#1}%
	\addtocounter{footnote}{-1}%
	\endgroup
}

\setlist[itemize]{leftmargin=*, nosep, align=parleft}
\setlist[enumerate]{leftmargin=*, nosep, align=parleft}

\AtBeginDocument{%
	\providecommand\BibTeX{{%
			\normalfont B\kern-0.5em{\scshape i\kern-0.25em b}\kern-0.8em\TeX}}}

\setcopyright{acmcopyright}
\copyrightyear{2018}
\acmYear{2018}
\acmDOI{10.1145/1122445.1122456}

\acmConference[Woodstock '18]{Public document}{June 03--05, 2020}{CSAIL, MIT}
\acmBooktitle{Neuroscience of program comprehension}
\acmPrice{15.00}
\acmISBN{978-1-4503-XXXX-X/18/06}

\begin{document}
	\title{Program Comprehension Does Not Primarily Rely On the Language Centers of the Human Brain}
    
    \author{Shashank Srikant\textsuperscript{1,2}, Anna A. Ivanova\textsuperscript{3}, Yotaro Sueoka\textsuperscript{3}, Hope H. Kean\textsuperscript{3}, Riva Dhamala \textsuperscript{4}, Evelina Fedorenko\textsuperscript{3}, Marina U. Bers\textsuperscript{4}, Un{a-M}ay {O'R}eilly\textsuperscript{1,2}\\
    {\small\texttt{\{shash, annaiv\}@mit.edu, evelina9@mit.edu, unamay@csail.mit.edu} }}
    \affiliation{\institution{${}^1$CSAIL, MIT \quad ${}^2$MIT-IBM Watson AI Lab  \quad ${}^3$BCS \& McGovern Institute for Brain Research, MIT,  \quad ${}^4$Eliot Pearson Dept. of Child Study and Human Development, Tufts University}\country{~}}

	\renewcommand{\shortauthors}{Shashank Srikant, \etal}

	\begin{abstract}
	Write abstract here.
	\end{abstract}
	

	\begin{abstract}
		\input{abstract}
	\end{abstract}

\begin{teaserfigure}
    \begin{tcolorbox}[
	standard jigsaw,
	opacityback=0]
    This manuscript is a rewrite of \textbf{Ivanova, A.A., Srikant, S., Sueoka, Y., Kean, H.H., Dhamala, R., O'Reilly, U.M., Bers, M.U. and Fedorenko, E., 2020. \textit{Comprehension of computer code relies primarily on domain-general executive brain regions.} eLife, 9, p.e58906} \cite{ivanova2020comprehension}.
    This manuscript informs the results of the eLife work to a computer science audience; the original eLife work was written to primarily inform the cognitive neuroscience community.
    \end{tcolorbox}
\end{teaserfigure}
	
	\maketitle

\input{introduction}
\input{relatedwork}

\input{background}
\input{design}

\input{procedure}
\input{results}
\input{discussion}
\input{threats}
\input{conclusion}
\newpage
\bibliographystyle{ACM-Reference-Format}
\bibliography{neuro-cs}
\newpage
\appendix
\input{method}

\end{document}

%% file: abstract.tex
%We understand little of the cognitive and neural bases of how we comprehend computer programs.
Our goal is to identify brain regions involved in comprehending computer programs. 
We use functional magnetic resonance imaging (fMRI) to investigate two candidate systems of brain regions which may support this -- the Multiple Demand (\md) system, known to respond to a range of cognitively demanding tasks, and the \lrspace, known to primarily respond to language stimuli.
We devise experiment conditions to isolate the act of \codecomp, and employ a state-of-the-art method to locate brain systems of interest.
We administer these experiments in Python (24 participants) and \scratch\xspace (19 participants) - which provides a visual interface to programming, thus eliminating the effect of text in code comprehension.
From this robust experiment setup, we find that the \lrspace is not consistently involved in \codecompi, while the \mdspace is.
Further, we find no other brain regions beyond those in the \mdspace to be responsive to code.
We also find that variable names, the control flow used in the program, and the types of operations performed do not affect brain responses.
We discuss the implications of our findings on the software engineering and CS education communities.

%% file: introduction.tex
\section{Introduction}
Reading and understanding computer programs (code) has been estimated to consume nearly $60\%$ of a software professional's time \cite{xia2017measuring}.
Yet, we understand little of how we cognitively accomplish it, making this an open question in science.
Extending seminal precedents \cite{siegmund, floyd}, we attempt in this work to study and establish the regions of the brain that are involved in comprehending computer code.

The recency of \codecompi as a cognitive skill suggests that brain regions which specialize in supporting other cognitive activities likely also support \codecompi.
Given its association with logic and problem solving, \codecompi can arguably be handled by regions responsible for working memory and cognitive control, or those involved in math and logic.
Similarly, code and natural language share many common properties. 
They possess similar syntactic and semantic structures, and hierarchically compose to convey meaningful information -- in both code and text, tokens are associated to form statements, which are further associated to form an entire code or document, which results in meaning being associated with the artifact \cite{fedorenko2019language}.
Arguably, regions of the brain involved in processing language can support \codecompi.

Neuroimaging research is well positioned to address which regions are involved in \codecompi. 
Techniques such as functional magnetic resonance imaging (fMRI) measure brain activity when performing cognitive tasks like reading or hearing music.
Brain regions whose functions have been well established, like language or music centers, responding to a new task, like \codecompi, can indicate the cognitive processes likely associated with that task \cite{mather2013fmri}.

In this work, we use fMRI to study how code-reading related tasks engage two known systems of brain regions -- the Multiple Demand (MD) and Language systems (details in Section \ref{background}).
While previous neuroimaging studies have also investigated brain regions involved in \codecompi, their results remain inconclusive. 
They provide evidence for activity in regions that roughly correspond to the \mdspace \cite{floyd, huang2019distilling, siegmund, siegmund2017measuring,  liu2020computer}, as well as in regions resembling the \lrspace \cite{siegmund, siegmund2017measuring}.
Importantly, these studies do not distinguish the act of \codecompi from other code-reading related activities like mentally simulating code.
Further, most do not quantify brain responses, and compare them to responses to other tasks associated with working memory or language to meaningfully interpret their observations.
We review these works in Sections \ref{sec:relatedwork} and contrast their design choices to ours in Section \ref{design}.

Our contributions in this work are twofold. 
First, we design novel experiments and introduce improved methods to identify brain regions involved in \codecompi.
Second, we present a new set of results which adds to our current understanding of the cognitive bases of \codecompi.
We summarize our design and method contributions below. 
See Section \ref{sec:results} for details on our results.
\begin{itemize}
	\item We offer a clearer definition of \codecompi, and design experiment conditions to isolate and measure it.
	\item We use a state-of-the-art procedure to determine which known, well-characterized brain systems respond to \codecompi.
	\item We test our experiments in two programming languages - Python and \scratch, a programming system with a fully visual interface, on a group of 24 and 19 participants respectively.
	Using \scratch enables measuring the effect of text on \codecompi, and additionally helps validate the generalizability of our results.
	Prior studies have experimented only with one programming language.
	\item We ensure that the observations we make generalize to different code properties like control flow (sequential programs, loops, conditionals), or types of operations performed (string, math operations).
	\item We additionally investigate whether brain activity corresponding to code in the \lrspace is a result of descriptive variable names used in codes.
	\item We make our code, stimuli, and brain data publicly available for the community to reuse and extend. Link - {\small\url{https://github.com/ALFA-group/neural-program-comprehension}}
\end{itemize}

%% file: relatedwork.tex
\section{Related Work}
\label{sec:relatedwork}

The question of whether there exist specialized regions in the human brain which are exclusive to specific cognitive functions goes back to Paul Broca's investigations of language understanding in the 1850s \cite{henderson1986paul}.
Advances in technology to accurately measure neural activity in the last three decades have revealed the existence of specialized regions for a variety of cognitive functions like language processing, face recognition, navigation \etc \cite{kanwisher2010functional}

The use of neuroimaging techniques to study the cognitive responses to programming has gained momentum recently.
Prior works have investigated the neural processes involved in debugging \cite{castelhano2019role}, variable tracking when reading programs \cite{ikutani2014brain, nakagawa2014quantifying}, semantic cues or program layout \cite{fakhoury2018effect, schroter2017comprehending}, program generation \cite{krueger2020neurological}, manipulating data structures \cite{huang2019distilling},  biases in code review processes \cite{huang2020biases}, and programming expertise \cite{floyd,parnin2017nature,ikutani2020expert}.

Relevant to our scope are works which investigate regions of the brain involved in comprehending code (as opposed to writing code, or any other coding-related activity).

\siegmund \cite{siegmund}, an influential work which pointed the community's attention to this topic, investigate the question of which regions in the brain are involved in code comprehension.
They present two sets of stimuli to 17 participants in an fMRI study. 
The first requires participants to read through snippets of code and determine their outputs. 
The second requires them to read code snippets with syntax errors and suggest fixes. 
The authors contrast activations from these two sets of stimuli, both of which correspond to code comprehension activity in the brain, to a baseline of no activity.
They show parts of this contrast to lie in the Broca's region (language centers) as defined by Brodmann's areas \cite{brodmann1909vergleichende}.

\floyd \cite{floyd} pose a different primary research question.
They investigate, on a larger sample of 29 participants, whether it is possible to distinguish the act of program comprehension from English sentence comprehension using brain activity measurements.
Their decoding experiments show that neural representations for code are unique and different from language.
As a secondary result, they do comment on brain regions involved, and partially confirm \siegmund's findings.
In their design, they use a baseline contrast of an English comprehension task and two code-reading tasks.
%In the second, participants review Git merge comments to either approve or reject them. 
%Additionally, they studied the role of expertise in comprehension.

\liu \cite{liu2020computer} very recently showed that code comprehension has very low overlap with the language centers of the brain, in line with the results we present in this work.
They present 17 expert programmers with two code-related tasks - the first is similar to \siegmund, where participants determine code output. 
The second requires participants to memorize what they call `fake code' -- code snippets with scrambled tokens in each line -- and confirm the presence of a specific substring.
They further administer math, logic, and language tasks to locate brain regions involved in these functions in every participant.
They report an overlap of code activity with regions belonging to the \mdspace but not the language centers.
%Their experiment design and analysis procedure makes it infeasible to clearly identify brain regions involved in code comprehension.

We pose the same question that \siegmund and \liu study.
We differ though in our experiment design and workflow. 
We compare our design choices to these works in detail in Section \ref{design}.
We shall see that this leads to a different set of conclusions than those of \siegmund

%% file: background.tex
\section{Background}
\label{background}

We provide a brief background on fMRI studies, what is measured by such scanning machines, and the regions of the brain we investigate.
\subsection{fMRI studies}
Functional magnetic resonance imaging (fMRI) is typically used to identify regions of the brain which respond to any cognitive task (comprehending code, in our case).
MRI machines can mark out and show brain responses in the order of a million voxels while sampling every few seconds \cite{glover2011overview}.
A voxel is roughly the 3-dimensional equivalent of a pixel, and spans a few cubic millimeters of our brains.
%Other methods like electroencephalography (EEG) and intra-cranial recordings offer varying spatial resolutions and sampling rates of brain responses.
 
When a brain region is involved in a cognitive task, blood flows into the region to aid its processing.
An MRI machine measures this change in blood-flow, and reports BOLD (blood oxygen level dependent) values sampled at the machine's frequency.
Following common practice, the parameters of a general linear model, fit to these time-varying values, are used as a metric for brain activity.
We provide details in Section \ref{procedure}.

\subsection{Regions of Interest (ROIs)}
We investigate whether two well-studied systems of brain regions -- the \mdspace and the \lrspace, which we know how to locate, are also activated when we comprehend code.
A \textit{region} (also referred to as \textit{parcel}) here denotes a contiguous chunk of brain mass involved in a cognitive task.
A \textit{system} of regions (also referred to as a \textit{network}) can comprise multiple disjoint (at the cortical level) regions, all involved in the same cognitive task.

\begin{figure}[hbpt]
	\begin{center}
		\includegraphics[width=0.9\linewidth ]{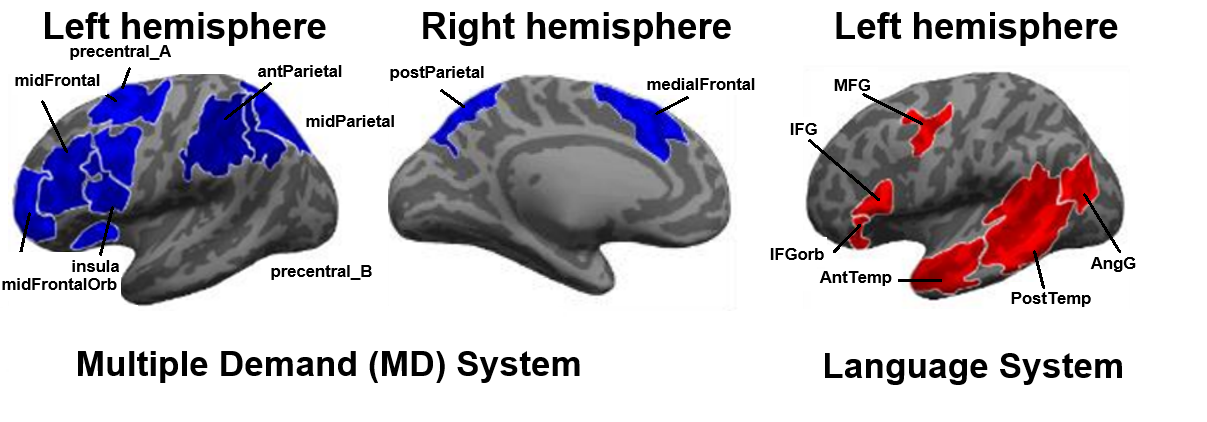}
		\caption{The Multiple demand  (\md) system and \lrspace highlighted in a neurotypical adult brain. 
		These two systems span multiple, closely situated regions in the brain, and have been established to have very different response profiles.
		What is conventionally referred to as Broca's region includes portions of both these systems \cite{fedorenko2020broca}.
		}
	\label{mdlr}
	\end{center}
\end{figure}

\par\textbf{Multiple Demand (\md) system.} 
Since programming conceivably involves arithmetic and general logic skills, we investigate whether the Multiple Demand system \cite{duncan2010multiple}, the most prominent system known to support these skills, is activated. 
Generally located in the prefrontal and parietal areas of the brain, this system of regions is known to be domain-agnostic, and is activated in a host of tasks requiring working memory and general problem solving skills, including math and logic \cite{duncan2010multiple, amalric2019distinct}.

\par\textbf{\lrspace.} 
Another possible candidate for processing code is the \lrspace.
These regions have been identified to respond to both comprehension and production of language across modalities (written, speech, sign language), respond to typologically diverse languages ($>50$ languages, from across $10$ language families), form a functionally integrated system, reliably and robustly track linguistic stimuli, and have been shown to be causally important for language \cite{fedorenko2010new, clark2003aphasia, blank2017domain, shainetal19, mineroff2018robust, blank2014functional}.

Figure \ref{mdlr} shows approximate locations of these systems in a neurotypical adult brain.
These systems have been consistently located roughly in the same parts of the brain across individuals \cite{fedorenko2010new, fedorenko2013broad}.
While an ROI provides a set of broad regions observed to be involved in a cognitive task across individuals, we further locate \textit{functional} ROIs (fROIs) -- specific voxels within these broad regions which respond to working memory and language respectively \textit{in an individual}.
By doing this, we account for the exact anatomical locations of these voxels, which vary across individuals.
This is one improved aspect of our experiment method over prior works.
We provide details on fROIs in Section \ref{design:analyze}.

%% file: design.tex
\section{Experiment Design}
\label{design}

We first provide a summary of our overall workflow. 
We follow that with details on three key components of our experiment design: \textbf{condition design} - the various design choices we consider in creating the code stimuli we show our participants, \textbf{fMRI tasks} - the tasks participants respond to in an MRI machine which enable measuring brain activities, and \textbf{processing fMRI data} - how we analyze participants' fMRI data and quantify the effect of \codecomp.
In our description of these components, we also contrast how they differ from previous works.

\subsection{Experiment workflow - An overview}
\label{expt-workflow}
The first step of our workflow is to frame hypotheses and design conditions which can test those hypotheses.
These conditions inform the stimuli and tasks we present to human participants in an MRI machine.
Our goal is to observe the effect reading code has on two regions of interest in our brains - the \mdspace and the \lrspace. 
We first determine which voxels (fROIs) belong to the \lrmdspace in each participant.
We do this by getting participants to respond to \textit{localizer tasks} -- tasks which have been shown to consistently activate the two systems \cite{fedorenko2010new, blank2014functional}.
We then show participants stimuli corresponding to our own carefully designed code conditions, and we measure brain responses to these conditions within the identified fROIs.
The goal of analyzing fMRI responses to our code conditions is to evaluate whether they activate the fROIs as much as the localizer tasks.
If they do activate the regions as much, we infer that the fROIs are involved in processing code.
For example, if comprehending code activates the \lrspace as much as comprehending English text (the localizer task for the \lrspace), we then conclude that the \lrspace is involved in processing code comprehension in addition to processing language comprehension.

%Critical to the success of such brain activity measurements are the design of the different conditions to contrast, and the analysis of corresponding fMRI signals.
%We now describe these steps in the following sub-sections.

\begin{figure*}[t]
	\begin{center}
		\includegraphics[width=0.9\linewidth ]{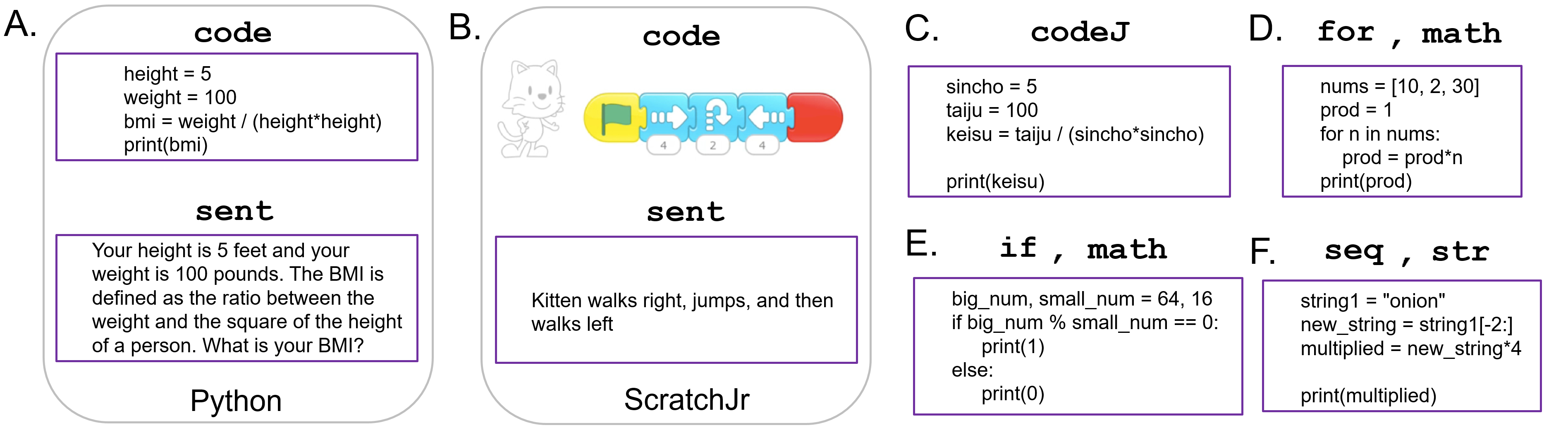}
		\caption{(A) A \texttt{code} condition stimulus in Python and its equivalent \texttt{sent} condition, which describes the \texttt{code} stimulus in words. 
		\texttt{sent} controls for brain responses to \codecontent.
		The difference in these conditions, \texttt{code}$>$\texttt{sent}, estimates \codecomp.
		(B) An example \texttt{code} and \texttt{sent} stimulus in \scratch, a programming system with a visual interface. 
		\scratch allows to measure the effect of text in codes.
		(C) \texttt{codeJ} condition with Japanese variable names, which controls for the effect of meaningful variable names.
		(D, E, F) Conditions that measure the effect of control-flow properties (\texttt{for, if, seq}) and type of operations (\texttt{math, str}).
		}
		\label{stimuli}
	\end{center}
\end{figure*}

\subsection{Condition design}
\label{design:conditiondesign}
Brain activity measurements for a given condition (\eg response to reading codes) can meaningfully be interpreted only relative to another condition (\eg response to reading plain text), \ie by \textit{contrasting} two or more conditions.
We describe the different conditions we design and contrast in our work, and discuss them in light of the design choices made by prior works.

\subsec{Controlling for \texttt{non-codes}.} The simplest condition pair to observe the effect of reading code is by contrasting \texttt{codes} with \texttt{non-codes} (notated as \texttt{code $>$ non-code} in the cognitive neuroscience literature). 
Here, \texttt{non-codes} correspond to stimuli which participants can comprehend despite not being code-like.
In our study, they correspond to statements in a natural language (English).
%We hence did not use this weaker condition.

Our goal though is to push farther.
We design conditions which help isolate the effect of other factors which might alternatively explain the activations we observe in different brain regions when understanding code.

%\begin{itemize}
\subsec{Controlling for \codecontent.}
Arguably, the task of reading code involves more than the act of \codecomp.
To appreciate why, consider the different cognitive steps involved in reading and understanding code. 
On being presented code -- 1) Retinal cells are activated by the presence of characters in a program
2) The visual system of our brain processes these characters.
3) Having recognized the characters, our brain interprets tokens present in the text.
4) Our brain groups tokens to recognize program statements, and eventually groups these statements to form a mental representation of the entire code, and understands its goal.
5) Our brain executes or simulates it to derive its final output.

In our work, we do not study the effects of reading code on the visual system (steps 1-2).
We identify steps 3-4 as \codecomp, and step 5 as carrying out \codecontent -- which has also been referred to as \textit{program tracing} \cite{soloway1986learning}, and \textit{processing code content} \cite{ivanova2020comprehension}.
For example, comprehending the statement \texttt{x=10+20} refers to associating this statement with the notion `\texttt{x} stores the sum of numbers $10$ and $20$'.
\codecontentcaps in this case refers to mentally adding numbers $10$ and $20$ and realizing that \texttt{x} stores $30$.
We refer to steps 3-5 collectively as \pu.

Step 5 can potentially dominate brain measurements made when reading and understanding code.
To factor out its effect, we offer the following insight -- it is possible to describe code in different ways while retaining its \codecontent operations.
A code described in sentences or as a flow diagram does not alter its operations.
%Regardless of whether one reads a program in its text form (tokens, lines of code) or as sentences describing the algorithm, the \textit{mental executions} of the code made to arrive at the final output remains the the same. 
Drawing on this insight, we design sentences whose content matches our code conditions.
We notate this condition as \texttt{sent} and the contrast as \texttt{code $>$ sent}.
See Figure \ref{stimuli}.A. for an example.
If the \texttt{code} condition measures \codecomp and \codecontent as the dominant cognitive steps involved, the \texttt{sent} condition then arguably measures natural language (sentences) processing and \codecontent.
The difference in these two conditions \texttt{code $>$ sent} thus allows us to isolate and measure the act of \codecomp.

\subsec{Controlling for variable names.} 
If \codecomp is indeed treated like language comprehension and the \lrspace is found to respond to it, it is reasonable to question whether the \lrspace responses are caused just by the presence of meaningful variable names and not other aspects of the code.
We control for this possibility by replacing variable names with those which mean nothing in that context.
The responses to such codes can then be attributed solely to \codecomp and not to the presence of meaningful English words in the code.
In our work, we chose to rename variables with their Japanese equivalent names (written out in the English script) and administer it to participants with no knowledge of Japanese.
We refer to this condition as \codej. 
Figure \ref{stimuli}.B shows the code in Figure \ref{stimuli}.A instead with Japanese variable names.
We also account for the effect meaningful string literals (\eg \texttt{x="hello"}) or meaningful keywords (\texttt{for}, \texttt{if}) may have, by designing an equal number of stimuli without these artifacts (discussed in the following point).

\subsec{Effect of control flow and operations.}
%We additionally investigate whether there exist brain regions sensitive to core properties of code. 
%We test this with the possibility of finding regions similar to those discovered by Hubel and Wiesel \cite{hubel1968receptive}, who found regions in the lower visual system of our brains that specialize in detecting and responding to the building blocks of vision like lines and edges.
We additionally investigate whether brain activations to code are consistent across different code properties.
This helps demonstrate the robustness of our observations to common variations possible in code.
We test two such properties -- control flow, and the types of operations.
In control flow, we test each of loops (\texttt{for}), conditional statements (\texttt{if}), and sequential statements.
See Figures \ref{stimuli}.D, E, F for examples of each of these conditions.
We test two types of operations -- math and string.
Figures \ref{stimuli}.E, F show examples of \texttt{math} and {str} operations respectively.
Every stimulus in these conditions has exactly one each of the three control structures, and one of the two data operations. 
This design also accounts for the presence of meaningful string literals and keywords by allowing us to observe brain activity corresponding to conditions that do not contain these artifacts (\texttt{math}, \texttt{seq} respectively).

\subsec{Effect of text in codes.} We experiment with the conditions we describe above in two programming languages -- Python and \scratch.
\scratch is a programming system with a fully visual interface \cite{bers2018coding}. 
It is generally introduced to children as means to express themselves creatively, where the visual interface and intuitive drag-and-drop features representing different programming constructs enable them to code without relying on a language like English \cite{bers2019coding}.
The very nature of this visual interface allows us rule out the influence of text on \codecomp.
Figure \ref{stimuli}.B shows an example.
Further, using \scratch as a second programming language helps validate the generalizability of our findings. 
All prior works have evaluated their findings only in one programming language.

\subsec{Design choices by prior works.} \floyd also use the basic contrast \texttt{code $>$ non-code}, but nothing more to isolate \codecomp.
\siegmund instead contrast \texttt{code $>$ code with syntax errors} (Section \ref{sec:relatedwork}). 
Codes with syntax errors are still codes, and hence do not help differentiate activity in regions where non-codes (natural language) are known to be processed.
Further, the \texttt{code-with-syntax-errors} condition likely measures aspects of \codecomp, \codecontent, and perhaps other skills specific to debugging and finding such errors.
Thus, their contrast does not fully isolate \codecomp.
While \liu ensure their \texttt{code} stimuli generalize to loops and conditions, their primary contrast \texttt{code} $>$ \texttt{fake code} also does not distinguish between \codecomp and \codecontent.
Their setup introduces the additional effect of memorizing \texttt{fake code} which involves multiple cognitive processes
%, but arguably involves neither \codecomp nor \codecontent.
%In addition to these crucial differences, these works do not consider other confounds we account for through our conditions.

\subsec{Summary.} To summarize, in our Python experiments, our overall experiment design is a $3\times3\times2$ study -- $3$ conditions - \texttt{code}, \texttt{sent}, \texttt{codeJ} (Japanese variable names), and within each of these three conditions, we further have $3$ categories of control flow conditions, and $2$ categories of operations-related conditions.
Since many of these conditions are not applicable to \scratch (variable names, operation types), we evaluate only the critical \texttt{code > sent} condition in \scratch.
%We create batches of tasks, each containing stimuli belonging to different conditions, and present them to participants in an MRI machine.
%For the primary code conditions \texttt{code} and \texttt{sent} (and \texttt{codeJ} in Python), every code problem has a version corresponding to each of these conditions.
%We provide details of these tasks and stimuli in Section \ref{procedure}.

\subsection{fMRI tasks}
\label{design:tasks}
For each participant, in addition to presenting stimuli corresponding to code-related conditions in an MRI machine, we present two separate tasks to \textit{localize} the two regions of interests in them.
What is central to a localizer task is its ability to strongly activate a region of interest in every individual.
It has been empirically established that reading semantically well-formed sentences in any natural language strongly activates the \lrspace, while performing spatial memory tasks strongly and distinctly activates the \mdspace \cite{fedorenko2013broad, fedorenko2010new}.
We reuse these established localizer tasks in our work.
We provide details in Section \ref{procedure}.

We now describe how we use this localization information when analyzing brain activity during \codecompi.
%\cite{fedorenko2009neuroimaging, fedorenko2012language}.\shash{confirm the right ref here}

\subsection{Locating fROIs and data analysis}
\label{design:analyze}
\input{gss}

%% file: gss.tex
We analyze brain data in the following five key steps.
Our procedure follows the Group-constrained Subject-Specific (GSS) method of locating functional regions of interest (fROIs) that are activated consistently across individuals \cite{nieto2012subject}.

\subsec{1. Mapping to an exemplar brain structure.} To normalize differences in brain anatomies, each participant's brain is spatially transformed to an exemplar brain structure like the Montreal Neurological Institute (MNI) template \cite{mni}.
These spatially transformed coordinates are used for subsequent analyses.

\subsec{2. Selecting ROIs.} 
Regions of interest (ROIs) mark out a set of broad regions observed to be involved in a cognitive task across individuals.
For every participant, we use these regions as a starting point, and look for voxels within them which respond to a cognitive task.
This helps avoid looking in regions which are not germane to the task.
For example, reading code will understandably also activate the visual cortex, which is not of interest to our particular study.

In our work, we reuse a set of 20 MD parcels (10 in each hemisphere) and six Language parcels defined in prior works \cite{fedorenko2010new, fedorenko2013broad}.
These parcels have been curated by aggregating  $\sim$200 participants' brain responses to spatial working memory and language tasks respectively.
As an alternate, one could select ROIs from the parcels defined by Brodmann's areas \cite{brodmann1909vergleichende}, an atlas which maps regions of an exemplar's brain to cognitive functions.

\subsec{3. Identifying fROIs.}
For every individual, a \textit{functional} region of interest (fROI) refers to a collection of voxels within an ROI which respond to the cognitive task the ROI is involved in.
%While ROIs are approximate locations of regions which have been identified across individuals, fROIs are importantly specific to individuals.
Owing to differences in anatomies, the specific set of voxels which respond to a cognitive task (like spatial reasoning or language) varies across individuals.
ROIs, aggregated from across individuals, help narrow down the search space to locate these specific voxels in every individual by pointing to a swath of regions known to respond to the task.
fROIs in turn identify specific voxels \textit{functionally involved} in the cognitive task.
Localizer tasks (Section \ref{design:tasks}) for each system help identify these voxels.
By the end of this step, we establish in each participant fROIs for the \lrmdspace.
%Prior works have shown brain activity measured from these individual-specific voxels to yield better estimates of regions, especially in the estimation of the \lrmdspace owing to their anatomical proximity \cite{fedorenko2009neuroimaging,fedorenko2012language}.
We provide details in Appendix \ref{mdlr}.

\subsec{4. Aggregate activation data within a participant.} 
We use the fROIs defined for the two systems in the previous step in all our remaining experiment conditions.
Specifically, we measure the activations of our code conditions in the selected fROIs.
At this stage, we have at least two sets of activation measurements for each voxel in an fROI -- one corresponding to the localizer task, and the others corresponding to the different code-related conditions.
For each fROI, we obtain a single response value per condition by averaging the responses of all voxels within the fROI.

\subsec{5. Aggregate activation data across participants.}
%We obtain group-level activation measurements by averaging the participant-level data obtained from step 4. above.
For each system, we then evaluate whether the distribution of participant-level responses to the code conditions is comparable to that of the localizer task. 
If it is, we conclude that the system is involved in processing code conditions.

\subsec{Multi-participant analysis without functional localizers.} 
Among prior works, \liu alone use localizer tasks to find task-selective voxels in individual participants.
However, they do not use ROIs (step 2 above) and instead perform a whole-brain analysis, and report overlaps as against measuring exact activations in fROIs.
Their setup coupled with their ambiguous condition design (discussed in Section \ref{design:conditiondesign}) makes it hard to infer brain regions accurately.

In fMRI studies which do not use localizer information, as in the case of \siegmund, \floyd, and other works which have studied different aspects of programming, the primary difference is that ROIs are defined based on anatomy, and not on their function (\ie how they respond to localizer tasks). 
Concretely, this difference arises in steps 3 and 4, where instead of aggregating activations within an fROI, activations are estimated in each voxel across the entire brain and aggregated across participants (also called the \textit{group analysis} procedure). 
The location of such aggregated active voxels is then described using anatomical labels, such as Brodmann areas \cite{brodmann1909vergleichende}.
This method has broadly been referred to as \textit{reverse inference} in neuroimaging studies \cite{poldrack2011inferring}.

The reverse inference method assumes that fROIs are spatially fixed among individuals and can be uniquely located in the exemplar brain structure.
While reverse inference is not always a concern, especially when the regions are anatomically well separated and distinct (\eg visual system vs. \mdspace), it has been shown to yield inaccurate estimates in the measurements of the closely situated \lrmdspace \cite{brett2002problem, amunts2012architecture, fedorenko2012language, fedorenko2020broca}.
What is referred to as the language region by Brodmann's areas (areas 44 and 45) in one  individual can instead refer to functional regions belonging to the \mdspace in another individual, owing to differences in individual anatomies \cite{fedorenko2012language, fedorenko2011functional}.
The GSS approach of function-based ROI identification helps circumvent this potential cause for inaccuracy.

%% file: procedure.tex
\section{Experiment Procedure}
\label{procedure}

We describe in brief our experiment procedure.
We provide details in Appendix \ref{method}.

We recruited $24$ participants for Experiment 1 (Python) and $19$ participants for Experiment 2 (\scratch), with no overlap between these groups.
On the day of the scan, having provided consent, participants spent $1.5-2$ hours in the scanner.
In Experiment 1, in the week of their scheduled fMRI scan, each participant additionally completed an assessment in Python to evaluate their fluency in it (Appendix \ref{method::participants}).

Once in the scanner, a participant was presented with two localizer tasks, adopted from prior works \cite{fedorenko2013broad, fedorenko2010new}, to locate the \mdspace and \lrspace respectively in their brain (Appendix \ref{method::locaizers}).
The \mdspace localizer task is a working memory task, presented in two grades of difficulty - easy and hard.
%The \lrspace localizer task requires reading sentences containing 12 words each (Appendix \ref{method::locaizers}).
The \lrspace task has two conditions - sentence reading (\sr), and non-word reading (\nr).
\sr requires reading sentences which are structurally and semantically meaningful.
%The example uses six words instead of 12 for brevity.
\nr requires reading sentences with pronounceable yet meaningless non-words (\eg\xspace {\small \texttt{BIZBY ACWORILLY BUSHU SNOOKI BILIBOP KUKEE}}).
These two conditions serve as references to measure other experiment conditions against -- the \lrspace has been shown to respond strongly to \sr while only minimally to \nr.

Participants were also presented with \putasks (Appendix \ref{method::pu}).
The tasks shown were balanced between the three conditions - \codee (code with semantically meaningful variable names in English), \sentt (sentences describing programs, controlled for \codecontent), and \codej (code with Japanese variable names).
Each participant saw 72 problems, 24 from each of the three conditions.
Each of these set of 24 problems further had an equal number of control-flow and operations-related conditions.
Any given participant saw only one of the three versions of a problem (Appendix \ref{method::procedure}).

The data from the localizer scans was used to locate the fROIs in the \lrmdspace in every participant (Appendix \ref{method::froi}).
We fit a general linear model to the time series brain activation data generated as a response to our different tasks.
The parameters of this model ($\beta$) are used as a metric for brain activity (BOLD) in all our analyses (Appendix \ref{method::dataprocess}).

\begin{figure*}[th]
	\centering
	\begin{minipage}{\textwidth}
		\makebox[\textwidth]{
			\resizebox{1.1\textwidth}{!}{
				\includegraphics[width=\linewidth]{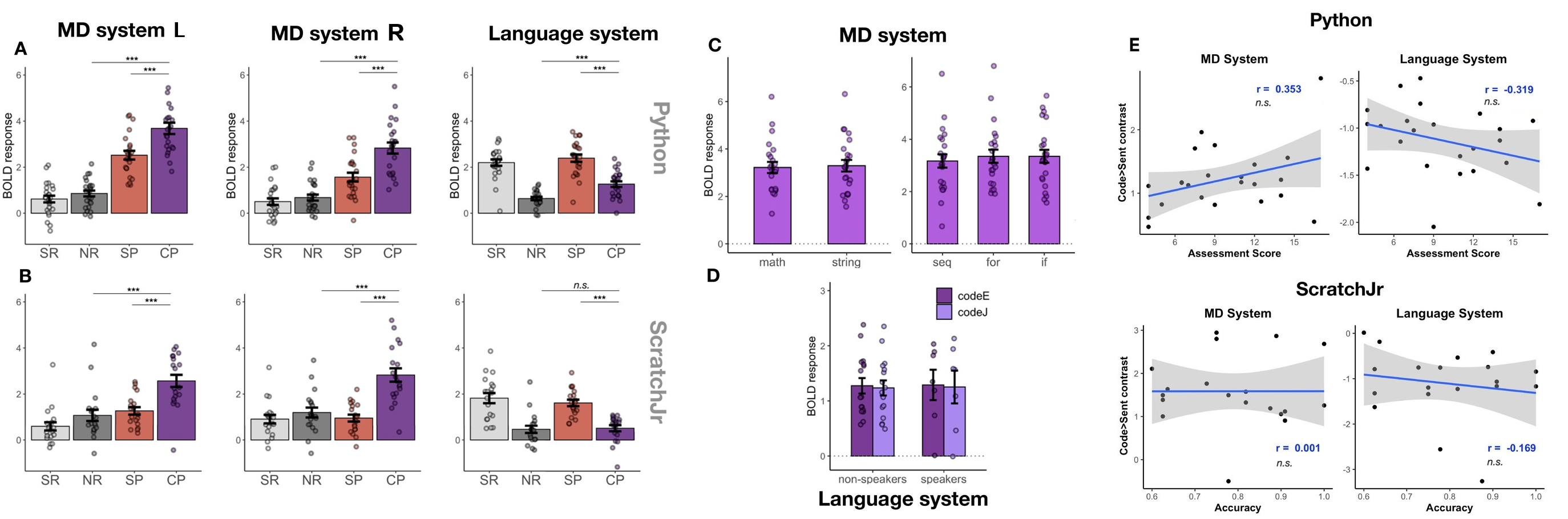}
			}
		}
	\end{minipage}
	\caption{(A, B) Brain activations in the \mdspace left hemisphere (\mdspace L), \mdspace right hemisphere (\mdspace R), and the \lrspace. 
		We measure responses to four conditions -- codes (\cp), sentences matching the code's operations (\sentp), Sentence reading (\sr), and Non-words reading (\nr).
		We experiment in Python (N=24) and \scratch (N=19).
		Each dot in the bars corresponds to aggregate data from one participant.
		\texttt{***} indicates $p<0.001$, \texttt{n.s.} - not significant 
		(C) \mdspace responses to two code properties -- operation type (math, string operations), and control-flow (sequential, loop (\texttt{for}), conditional (\texttt{if}))
		(D) \lrspace responses to variable names in English (\codee) and Japanese (\codej)
		(E) Correlation of responses in the \lrmdspace to proficiency in Python (top) and \scratch (bottom).}
	\label{results}
\end{figure*}

%% file: results.tex
\section{Results}
\label{sec:results}
%We study the effect of our \putasks on two candidate system of brain regions -- the \mdspace and the \lrspace.
%We study these effects in two languages - Python (Experiment 1) and \scratch (Experiment 2).
%We primarily investigate whether the \lrmdspace process code.
%Based on our findings, we supplement them with other research questions. 
We present our questions and their corresponding results here.
In our results, we discuss the neural activations in different regions of the brain (Figures \ref{results}.A, \ref{results}.B).
The x-axis in these plots corresponds to the different conditions participants responded to, and the y-axis represents activation strength ($\beta$ values, see Appendix \ref{method::dataprocess} for details).
Each dot in each bar corresponds to one participant's aggregate activity in the fROIs localized in them.
When reporting results of a contrast between any two conditions, we measure the difference in the average $\beta$ values (\delbeta) and compute its associated p-value.

%%%%%%%%%% RQ1
\begin{tcolorbox}[
	standard jigsaw,
	opacityback=0,  % this works only in combination with the key "standard jigsaw"
	]
	\textbf{\textsf{RQ 1.}} Does \pu activate the Multiple Demand (\md) system?
	
	\textbf{Conditions contrasted.} \texttt{code}, \texttt{sentence reading}, \texttt{non-word reading}
\end{tcolorbox}
We begin by investigating whether reading code, which involves both \codecomp and \codecontent, activates the \mdspace.
We do this by comparing the activations of our primary code-related condition - code problems, to the localizer tasks for the \lrspace\xspace - sentence reading and non-word reading.
We notate these conditions as \cp, \sr, and \nr respectively in Figure \ref{results}.A, B.
We evaluate two sets of fROIs in the \mdspace\xspace - one in each hemisphere of the brain (Figure \ref{results}.A, B., left and center plots).
From the plots, we see both sentence reading and non-word reading, the language localizer conditions, have minimal activations in the \mdspace in both hemispheres. 
This is expected since the \mdspace is not sensitive to language tasks \cite{blank2014functional}.
We find that code problems, which account for both \codecomp and \codecontent, activate fROIs in both hemispheres of the \mdspace consistently and significantly more than the baselines in both our experiments (Python: \delbeta$=2.17$, p $<0.001$; \scratch: \delbeta$=1.23$, p $<0.001$).
This suggests that the \mdspace is involved in reading code.

We confirm whether these responses are consistent across code properties, which will establish its robustness to the variations possible in code.
We test two properties -- control-flow (sequential, \texttt{for}, \texttt{if}), and types of data manipulated in them (string, math operations) in Python.
We observe strong responses regardless of the type of operations and control flow (Figure \ref{results}.C; y-axis is response to \cp).
We thus conclude that the responses in the \mdspace to code problems were not a result of any one particular type of problem, or mental operations related to a particular control flow.

This clearly identifies and establishes the role of the \mdspace in \pu.
Prior works did not identify and study this system of regions.
%only reported evidence of activity in these broad regions.

\begin{tcolorbox}[
	standard jigsaw,
	opacityback=0,  % this works only in combination with the key "standard jigsaw"
	]
	\textbf{\textsf{RQ 1 result.}} Yes, \pu activates the \mdspace.
	Its responses are independent of the control-flow operations and types of data operations present in codes.
\end{tcolorbox}

\begin{tcolorbox}[
	standard jigsaw,
	opacityback=0,  % this works only in combination with the key "standard jigsaw"
	]
	\textbf{\textsf{RQ 2.}} Does \codecomp activate the Multiple Demand (\md) system?
	
	\textbf{Conditions contrasted.} \texttt{code}, \texttt{sent}, \texttt{sentence reading}, \texttt{non-word reading}
\end{tcolorbox}
Since we find that \pu activates the \mdspace, we investigate whether these were responses to \codecomp or \codecontent.
To answer this, we study the effect of both our code-related conditions -- code problems (\cp), and sentence problems which match the code problems for their content (\sentp).
We find that sentence problems, which measure only \codecontent and not \codecomp, activate the \mdspace significantly greater than the language localizer baselines in both hemispheres only for Python (left: \delbeta$=1.51$, p $<0.001$; right: \delbeta$=0.78$, p $<0.001$). 
This activation is not significant for \scratch (left: \delbeta$=0.09$, p $=0.93$; right: \delbeta$=-0.40$, p $=0.004$), suggesting that \codecontent is not consistently supported by the \md.
However, we find that code problems, which measure both \codecomp and \codecontent, strongly activate fROIs in both hemispheres.
This is despite sentence problems taking slightly longer on average to respond to (Appendix \ref{beh_results}).
%Recall that the difference between the two conditions \cp and \sentp estimates \codecomp.
We hence find that \cp strongly activates the \mdspace and \sentp does not.
This implies that the difference \cp $>$ \sentp, which measures \codecomp, strongly activates it.
This is strong evidence for the \mdspace's consistent and robust activation to \codecomp, and shows it is not just a response to the underlying \codecontent operations.

We investigate further for any hemispheric bias towards \codecomp. 
Previous works have shown that math and logic problems typically activate the \mdspace in the left-hemisphere of the brain \cite{amalric2016origins, amalric2019distinct}.
We did not find any such bias in Python (\md-L plot, Figure \ref{results}.A). 
In \scratch, we observe stronger responses in the right hemisphere (\delbeta$=0.57$, p $<0.001$; \md-R plot, \ref{results}.B), perhaps reflecting a known bias of the right-hemisphere towards visuo-spatial processing \cite{sheremata2010hemispheric}.
%Using localizer tasks for math activity can help quantify this observation.

Follow up analyses of activity within individual regions within the \mdspace showed that $17$ of the $20$ fROIs in the Python experiment, and  $14$ of the $20$ fROIs in the \scratch experiment responded significantly more strongly to code problems than to sentence problems (details in Appendix \ref{froi_results}).
This demonstrates code processing is broadly distributed across the \mdspace and is not localized to a particular subset of regions within it.
Within this activated subset, we evaluate whether any fROIs are \textit{selective} to code problems in comparison to other cognitively demanding tasks which activate the \mdspace.
We find none for \scratch, and three regions in the frontal lobe (precentral-A, precentral-B, midFrontal) which exhibit stronger responses to Python code problems than to the hard working memory localizer task for the \mdspace. 
However, the magnitude of \texttt{code} $>$ \texttt{sent} in these regions (\delbeta$=1.03, 0.95, 0.97$) was comparable to the mean magnitude across all \mdspace fROIs (average \delbeta$=1.03$), suggesting that the high response was caused by the underlying \codecontent rather than \codecomp.
We conclude that \codecomp is broadly supported by the \mdspace, and no specific regions in the \mdspace are functionally specialized for it.

These new results further establish the role of the \mdspace in processing \codecomp, which we narrowly and clearly define in this work. 

\begin{tcolorbox}[
	standard jigsaw,
	opacityback=0,  % this works only in combination with the key "standard jigsaw"
	]
	\textbf{\textsf{RQ 2 result.}} Yes, \codecomp consistently activates the \mdspace. 
	Unlike math and logic, it activates fROIs in both the left and right hemispheres.
	In fact, no specific fROI within the \mdspace specializes for \codecomp, and it is instead broadly supported by the entire system.
\end{tcolorbox}

\begin{tcolorbox}[
	standard jigsaw,
	opacityback=0,  % this works only in combination with the key "standard jigsaw"
	]
	\textbf{\textsf{RQ 3.}} Does \pu activate the \lrspace?
	
	\textbf{Conditions contrasted.} \texttt{code}, \texttt{sent}, \texttt{sentence reading}, \texttt{non-word reading}
\end{tcolorbox}

We investigate the \lrspace similarly for responses to our code conditions.
Figures \ref{results}.A, B (rightmost plot) show the aggregate responses in the \lrspace to the two code conditions and the two language localizer conditions described above.
As expected of the localizers, we find the activations of sentence reading to be significantly greater than non-word reading \cite{blank2014functional, fedorenko2010new}.
Among the code conditions, we find that sentence problems activate the \lrspace as much as the sentence localizer task in both Python and \scratch. 
This is again expected since sentence problems contain English sentences describing what the program does (Figure \ref{stimuli}.A).
However, the responses to code problems were weaker than responses to sentence problems in both experiments (Python: \delbeta$=0.98$, p $<0.001$, \scratch: \delbeta$=0.99$, p $<0.001$).
This observation alone does not yield any insight on whether code activates the \lrspace, and we hence compare these activations to the localizer baseline non-word reading. 
Non-word reading is a lower bound for activity in the \lrspace; this is the activity seen in the \lrspace when it is not actively engaged in linguistic interpretation.
Responses to the code condition were stronger than non-word reading only in the Python experiment (\delbeta$=0.78$, p $<0.001$) but not in the \scratch experiment (\delbeta$=0.15$, p$=0.29$), implying that code does not consistently activate the \lrspace.
 
The result from this principled investigation of the \lrspace is contrary to that of \siegmund, who report the involvement of the language system in addition to other brain regions. 
We discuss this further in Section \ref{sec:discuss}.

\begin{tcolorbox}[
	standard jigsaw,
	opacityback=0,  % this works only in combination with the key "standard jigsaw"
	]
	\textbf{\textsf{RQ 3 result.}} No, \pu does not consistently activate the \lrspace.
\end{tcolorbox}
%%%%%%%%%% RQ4
\begin{tcolorbox}[
	standard jigsaw,
	opacityback=0,  % this works only in combination with the key "standard jigsaw"
	]
	\textbf{\textsf{RQ 4.}} Do meaningful variable names affect the \lrspace's response to code?
	
	\textbf{Conditions contrasted.} \texttt{codeE}, \texttt{codeJ}
\end{tcolorbox}
Since we find that Python code activates the \lrspace but \scratch does not, we investigate whether this is a consequence of meaningful variable names present in codes.
To study this effect, we had participants read half the Python code problems with semantically meaningful variable names in English (\codee) and the other half with Japanese variable names (\codej), making them semantically meaningless; 18 of the 24 participants reported no knowledge of Japanese.
%We observe activations only in the \lrspace since the \mdspace is not sensitive to words, regardless of whether they are meaningful or not.
In the \lrspace, we found no effect of meaningful variable names (\delbeta$=0.03$, p $=0.84$) (Figure \ref{results}.D, non-speakers), knowledge of Japanese (\delbeta$=0.03$, p $=0.93$) (Figure \ref{results}.D, speakers), nor any interaction between the two (\delbeta$=0.09$, p $=0.71$), suggesting that the \lrspace response was not affected by the presence of semantically meaningful variable names.
This result is surprising since the \lrspace has been shown to be very sensitive to word meaning \cite{anderson2019multiple}.
A possible explanation is that participants do not fully engage with the words' meanings to solve problems. 
%Another is that in the context of solving these problems, variables are assigned abstract rather than any concrete meaning.

\begin{tcolorbox}[
	standard jigsaw,
	opacityback=0,  % this works only in combination with the key "standard jigsaw"
	]
	\textbf{\textsf{RQ 4 result.}} Meaningful variable names do not affect the \lrspace's  response to code.
\end{tcolorbox}

%%%%%%%%%% RQ5
\begin{tcolorbox}[
	standard jigsaw,
	opacityback=0,  % this works only in combination with the key "standard jigsaw"
	]
		\textbf{\textsf{RQ 5.}} Are there regions outside the \mdspace and \lrspace that respond to \codecomp?
\end{tcolorbox}
To search for regions responsive to \codecomp outside the \mdspace and \lrspace, we perform a whole-brain Group-constrained Subject Specific analysis.
For both Python and \scratch, we search for brain areas with activations where \texttt{code} $>$ \texttt{sent}.
We then examine the response of such regions to code and sentence problems (cross-validated with held-out data), as well as to conditions from the two localizer experiments.
In both experiments, the discovered regions spatially resembled the \mdspace. 
For Python, any region that responded to code also responded to the spatial working memory task (\mdspace localizer). 
In case of \scratch, some fROIs responded more strongly to code problems than to the spatial working memory task; these fROIs were located in early visual areas/ventral visual stream which likely responded to low-level visual properties of \scratch code (which contains colorful icons, objects, \etc). 
%The traditional random-effects group analyses revealed a similar activation pattern.
These whole-brain analyses demonstrate that the \mdspace responds robustly and consistently to \codecomp, confirming the results of the fROI-based analyses discussed in RQs 1 and 3. 
This further shows that fROI-based analyses did not miss any non-visual regions outside the boundaries of the \lrmdspace that was activated by \codecomp.

\begin{tcolorbox}[
	standard jigsaw,
	opacityback=0,  % this works only in combination with the key "standard jigsaw"
	]
	\textbf{\textsf{RQ 5 result.}} We found no code-selective regions outside the \lrmdspace.
\end{tcolorbox}

%%%%%%%%%% RQ6
\begin{tcolorbox}[
	standard jigsaw,
	opacityback=0,  % this works only in combination with the key "standard jigsaw"
	]
	\textbf{\textsf{RQ 6.}} Does expertise affect how the \lrmdspace respond to \codecomp?
\end{tcolorbox}
%We analyze whether the activations to \codecomp in the \mdspace and \lrspace regions vary with the level of programming expertise. 
We study the role of expertise by correlating responses within each system with independently obtained proficiency scores for participants of Experiment 1 (a 1-hour Python assessment module; Appendix \ref{method::participants}) and with in-scanner accuracy scores for Experiment 2 participants.
Figure \ref{results}.E plots the percentage proficiency scores (x-axis) against \codecomp (\texttt{code} $>$ \texttt{sent}).
No correlations were significant. 
However, due to a relatively low number of participants (N = 24 and N = 19, respectively), these results should be interpreted with caution.
%
%\begin{figure}[hbpt]
%	\begin{center}
%		\includegraphics[width=0.9\linewidth ]{img/expertise.jpg}
%	\end{center}
%	\caption{Various results from our work}
%	\label{results2}
%\end{figure}

\begin{tcolorbox}[
	standard jigsaw,
	opacityback=0,  % this works only in combination with the key "standard jigsaw"
	]
	\textbf{\textsf{RQ 6 result.}} We did not have enough data to observe the effect of expertise on the \lrmdspace' responses to \codecomp.
\end{tcolorbox}

%% file: discussion.tex
\section{Discussion}
\label{sec:discuss}
We present a new set of results which improves our understanding of the cognitive bases of program comprehension.
We find that \pu (which we identify to include both \codecomp and \codecontent) strongly activates only the \mdspace and not the \lrspace.
Despite their anatomical proximity in the left-hemisphere of our brains, our work clearly distinguishes the roles of both these systems by means of functional localizers.

\subsec{\mdspace results.}
We find that the \mdspace consistently processes \pu.
We support our observations by showing these activations generalize across two code properties - control flow and data operations, suggesting that the system's response is robust to variations in code.
We further learn that the \mdspace responds consistently to \codecomp.
It also responds to \codecontent strongly in Python, but we see only a weak evidence for it in \scratch, which needs to be investigated in future work.
It is reasonable to expect the \mdspace to process \pu, since both \codecomp and \textit{simulation} requires attention, working memory, inhibitory control, planning, and general flexible relational reasoning - cognitive processes long linked to the \mdspace \cite{duncan2000common, duncan2010multiple}.
This finding also supports \liu's recent results \cite{liu2020computer}.
Since no other regions outside the \mdspace responded to codes, we posit this system stores code-specific knowledge in addition to processing it. 
This knowledge likely includes concepts specific to a programming language (\eg the syntax marking an array in Java vs. Python) and concepts shared across languages (loops, conditions). 
Evidence from studies on processing mathematics and physics \cite{fischer2016functional,amalric2019distinct} has shown that the \mdspace can store some domain-specific representations in the long term, perhaps for evolutionarily late-emerging and late-acquired domains of knowledge. 
In conclusion, we identify a known brain system, which had previously not been studied for its role in \pu tasks, to be involved in \codecomp specifically.

\subsec{\lrspace results.}
We importantly establish in this work that \pu is \textit{not} consistently processed by the \lrspace.
This is a new finding, and adds to the results from \siegmund and \floyd, while confirming results from \liu.
\siegmund report the involvement of the language centers in \codecomp by showing evidence of left-lateralized brain activity.
While it is unclear whether their observations were technically from the \lrspace or the \mdspace, we suspect that they observed \codecontent and not \codecomp.
Additionally, and surprisingly, we find that the \lrspace is insensitive to the presence of meaningful variable names.
More work is required to determine why the \lrspace showed some activity in response to Python code.

The \lrspace does not respond consistently to \codecomp despite numerous similarities between code and natural language. 
However, the lack of consistent \lrspace engagement in \codecomp does not mean that the mechanisms underlying language and code processing are completely different.
It is possible that both the \lrmdspace have similarly organized neural circuits that allow them to map a symbol to a concept. 
However, the fact that we observed code-related activity primarily in the \mdspace indicates that \codecomp does not activate the same neural circuits as language, and needs to use domain-general \mdspace circuits instead.

Having identified the regions activated when reading programs, we discuss how our results affect the programming languages (PL), software engineering (SE), and CS education (CS-Ed) communities.

\subsec{Impact on the PL, SE, CS-Ed communities.} To understand how our results can be applied specifically to \textit{improving} how we can understand programs, we first establish the relationship between two cognitive activities engaging the same brain system (in our case - working memory tasks and \codecomp engaging the \mdspace).
A few studies have claimed that for any two cognitive activities that share the same brain resources, training one activity will lead to an improvement in the other \cite{jaeggi2008improving, melby2013working}.
For example, if language and music share and activate the same brain system, then tools and approaches used to engage and train one activity should be transferable to, and will lead to an improvement in the other.
Since the effects of training and improving one's \mdspace are not well understood, it is unclear whether training on cognitively demanding non-coding tasks could improve our ability to read and understand programs. 
%That said, we recognize the general role the \mdspace may play in the ease of program comprehension -- more entities to keep note of and track in a program will arguably increase \mdspace activations which, as a general principle, one may want to avoid for better comprehension.

Based on the opposite conclusions presented in \siegmund, Portnoff \etal \cite{portnoff2018introductory} and similar other works suggest adopting a ``languages first" approach when teaching programming.
Evidence from our work does not support this claim, and we caution against adopting practices that are used to teach natural languages for programming just based on conclusions from recent neuroimaging studies.

The \lrspace not being involved in \codecomp should not diminish the role of language in understanding programs. 
The use of poorly named variables has been shown to increase cognitive load \cite{fakhoury2019measuring}, and non-native English speakers have been found to often struggle with learning English-based programming languages \cite{guo2018non}.
Future work should reconcile this disparity, and aim to show how results from studies on cognition can aid understanding programs.

%The trajectory involved in acquiring \codecomp skills is not well understood.
%Do other brain regions, perhaps even the \lrspace, take over \codecomp and \textit{simulation} after gaining expertise?
%It is possible that the \lrspace may play a role in learning to program \cite{prat2020relating}, even if it is not required to support code comprehension once the skill is learned. 
%How prior knowledge of a natural language affects our ability to read and understand programs needs to be studied further.
%Attempts have also been made to design programming languages that are closer to natural language (\eg Cobol, Python).
%While these studies establish a well understood connection between natural languages and program comprehension, our results do not directly influence this relationship. 
%The role of prior knowledge of a natural language and its effects on our ability to read and understand programs needs to be studied further.

\subsec{ML models for code.}
Recent advances in machine learning (ML) models trained on large corpora of programs have shown models' ability to perform tasks like renaming functions, bug detection, \etc \cite{allamanis2018survey}.
Deep networks learn a `language model' of programs, and likely learn a generalized way to represent these programs.
Do these model-learned representations correspond to the representations (activation data) in the different fROIs from our study?
Such a correspondence may have far reaching implications.
%If we do find the existence of such a correspondence, it will help isolate different aspects of what the model learns and confirm whether those aspects are learned and represented in our brains as well. 
For instance, if we can probe and isolate specific weights or layers that encode loops and recursion in a recurrent model like \textsf{seq2seq}, \textsf{code2seq}, or \textsf{GPT-3}, the existence of a correspondence between representations may help locate the encodings of loops and recursion in our brains.
Such a correspondence has recently been established between representations stored in the visual cortex and those learned by deep convolutional networks for image processing and recognition \cite{yamins2014performance, khaligh2014deep, cadena2019deep}. 
This promises to be a compelling direction for future work.

%% file: threats.tex
\section{Threats to validity}
There are limitations to the results we report in this work.
One possible threat is posed by the tasks we designed -- do our programming tasks measure code comprehension and understanding?
The programs we present in this study are short snippets of procedural code with limited program properties. 
They do not have complex control and data dependencies generally seen in production-grade programming projects.
Properties like function calls, objects, types, complex data, and state changes in the program are not included either and should be studied in the future, building on the understanding of simpler snippets provided by our work.
Further, we study a very specific activity related to programming -- reading programs, and do not investigate other equally important aspects like designing solutions, selecting appropriate data structures, and writing programs.

In designing our code stimuli, having overtly informative variables names poses the risk of participants not going through all the lines of code presented to them, and instead just guessing what the code does by gleaning variable names.
To avoid this, we constrain our variable names to be informative and natural (as it would appear in a real codebase) to the extent they do not reveal fully the intent of the entire code snippet.
However, such a constraint does not appear in actual coding scenarios.
Disparate stimuli are a source of possible confounds.
If some conditions had disproportionately longer code than others, it would be unclear if any trend we saw in brain activations were a function of the condition, or such factors like code length.
In an attempt at avoiding this, we ensured that the stimuli in our $3\times3\times2$ conditions had similar code lengths and overall response times. 
However, they are not all equal.

Expertise is also a potential confound which could affect the generalizability of the results we see. 
The majority of our participants were recruited from a university setting and had roughly 3-6 years of programming experience.
While our participants' experience level was largely homogeneous in our study, expertise could interact with brain functions associated with program comprehension, as it does with other cognitive functions \cite{agrawal2018does, hoenig2011neuroplasticity, gomez2019extensive}.
Accounting for the role of expertise would require running these experiments on a population with varying proficiencies.

Neuroimaging experiments generally risk lack of generalizability of results owing to low sample sizes.
In our work, the relatively small sample sizes (N=24 for Experiment 1, and N=19 for Experiment 2) affect only our group analyses (of comparing aggregate information across participants).
%The activation data from within each participant does not suffer from low data -- we aggregate informations across thousands of voxels within parcels of fROIs.
%In our group analyses, we compare our effects using statistical tests with low significance levels ($<0.001$).
Although these sample sizes are the norm in the neuroimaging community, the robustness of our results are limited by the small number of participants. 
We see this as an opportunity for authors from similar neuroimaging studies to collaborate to analyze data across these works, which will also help amortize the high costs of carrying out this type of experiments.

%We also note here that implications of our results is one-directional -- if we find code-related stimuli to activate an fROI at least as much as the localizer task for that fROI, then we conclude that the fROI is involved in code-related processing.
%We cannot however conclude whether the fROI is necessary for our ability to process code; such results only signify sufficiency.

%% file: conclusion.tex
\section{Conclusion}
Our work presents a new set of methods, experiments, and results which furthers our understanding of how our brains comprehend computer code. 
It is unique in addressing two core issues that made it difficult to interpret results from prior studies. 
First, we disentangle responses to the act of \codecomp from \codecontent.
Second, we analyze responses in brain areas that are functionally localized in individual participants, which provides an accurate interpretation of the observed responses.
We find that the \mdspace in our brains consistently processes \codecomp, while the \lrspace does not.
%Our findings suggest that earlier reports of left-lateralized activity from \siegmund \cite{siegmund} may reflect \codecontent rather than \codecomp. 
%This distinction between the two activities should be considered when interpreting results of other recent studies of programming effects on brain activity.
We release our code, stimuli, and brain data for the community to reuse and extend.
\balance

\section{Acknowledgements}
This research was partially supported by NSF grant 1744809. 
SS and UMO’R were supported by a grant from the MIT-IBM Watson AI Lab.

%% file: method.tex
\section{Methods}
\label{method}

In this section, we present details on the stimuli we used for the fROI localizer tasks, how we defined fROIs, stimuli we designed for the \putasks, the procedure we followed to present these tasks to our study participants, how we recruited our participants, and finally, our analysis of the recorded MRI data.

\subsection{Participants} 
\label{method::participants}
For Experiment 1, we recruited 25 participants (13 women, mean age = 22 years, SD = 3.0). 
Average age at which participants started to program was 16 years (SD = 2.6); average number of years spent programming was 6.3 (SD = 3.8). 
In addition to Python, 20 people also reported some knowledge of Java, 18 people reported knowledge of C/C++, 4 of functional languages, and 20 of numerical languages like Matlab and R. 
Twenty-three participants were right-handed, one was ambidextrous, and one was left-handed; the left-handed participant had a right-lateralized \lrspace and was excluded from the analyses, leaving 24 participants total (the rest had left-lateralized language regions). 
All participants in Experiment 1 were also provided a 1-hour paper-pencil Python assessment, administered in the week of their scheduled scan.
It consisted of 23 questions, 22 of which expected short-responses. 
One question required an algorithm to be designed and implemented in Python.
There was no overlap between questions on this assessment and the stimuli presented in the experiments.
We used this to measure their general proficiency in Python and programming.

For Experiment 2, we recruited 21 participants (13 women, mean age = 22 years, SD = 2.8). 
There was no overlap in participants between the two experiments.
In addition to \scratch, 8 people also reported some knowledge of Python, 6 people reported knowledge of Java, 9 people reported knowledge of C/C++, 1 of functional languages, and 14 of numerical languages like Matlab and R (one participant did not complete the programming questionnaire). 
Twenty were right-handed and one was ambidextrous; all participants had right-lateralized language regions, as evaluated with the language localizer task. 
Two participants from Experiment 2 had to be excluded due to excessive motion levels during the MRI scan, leaving 19 participants total. 

All participants were recruited from local universities and their surrounding communities, and compensated for their participation. 
All were native speakers of English, had normal or corrected to normal vision, and reported working knowledge of Python or \scratch, respectively. 
The protocol for these studies was approved by an institutional review board. 
All participants gave written informed consent in accordance with protocol requirements.

\subsection{Localizer tasks}
\label{method::locaizers}
All participants completed a language localizer task aimed at identifying language-responsive brain regions \cite{fedorenko2010new}, a spatial working memory localizer task aimed at identifying the multiple demand (MD) brain regions \cite{fedorenko2013broad}, and a set of \putasks.

A language localizer task identified brain regions within individual participants that selectively respond to language stimuli. 
During the task, participants read sentences (\eg~{\tiny NOBODY COULD HAVE PREDICTED THE EARTHQUAKE IN THIS PART OF THE COUNTRY}) and lists of disconnected, pronounceable non-words (\eg~{\tiny U BIZBY ACWORRILY MIDARALBUSHU SNOOKI BILIBOP KUKEE WEPS WIBRON PUZ}).
Each stimulus consisted a total of twelve words/non-words.
For details of how the language materials were constructed, see Fedorenko \etal \cite{fedorenko2010new}. 
We refer to the sentence reading task as \sr and non-word reading task as \nr.
The \sr $>$ \nr contrast has been previously shown to reliably activate left-lateralized fronto-temporal language processing regions and to be robust to changes in the materials, task, and modality of presentation \cite{fedorenko2010new, mineroff2018robust}. 
Stimuli were presented in the center of the screen, one word/non-word at a time, at the rate of 450 ms per word/non-word. 
Each stimulus was preceded by a 100 ms blank screen and followed by a 400 ms screen showing a picture of a finger pressing a button, and a blank screen for another 100 ms, for a total trial duration of 6 s. Participants were asked to press a button whenever they saw the picture of a finger pressing a button. 
This task was included to help participants stay alert and awake.

\begin{figure}[th]
	\begin{center}
		\includegraphics[width=0.9\linewidth ]{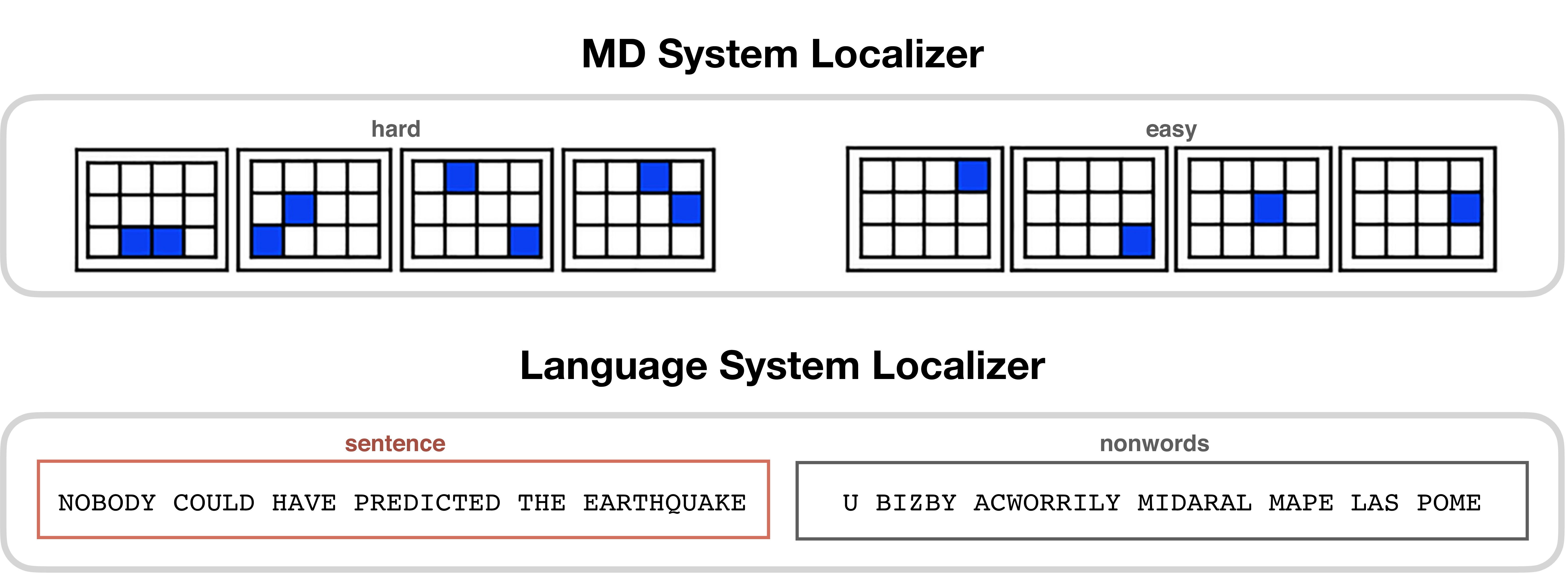}
		\caption{The two localizer tasks we adopted from prior works \cite{fedorenko2010new, fedorenko2013broad}. 
			The \mdspace localizer task tests spatial reasoning. 
			Participants are shown, in quick succession, 4 screens with different highlighted squares in a grid. 
			They are then shown two grids and have to correctly identify the grid which superimposes all the highlighted squares shown to them.
			The \lrspace localizer task requires participants to read two sets of sentences - one coherent and meaningful (left), and the other pronounceable yet meaningless (right).
		}
		\label{localizer_appendix}
	\end{center}
\end{figure}

Previous work established a spatial working memory task which identified the MD system in individuals \cite{fedorenko2013broad}.
Participants had to keep track of four (the easy condition) or eight (the hard condition) sequentially presented locations in a 3$\times$4 grid (see Figure \ref{stimuli}). 
In both conditions, they performed a two-alternative forced-choice task at the end of each trial to indicate the set of locations they just saw. This \texttt{hard $>$ easy} contrast has been previously shown to robustly activate MD regions \cite{fedorenko2013broad}.
These regions have been shown to respond to difficulty manipulations across many diverse tasks \cite{duncan2010multiple, fedorenko2013broad}. 
Stimuli were presented in the center of the screen across four steps. Each of these steps lasted for 1000 ms and presented one location on the grid in the easy condition, and two locations in the hard condition. 
Each stimulus was followed by a choice-selection step, which showed two grids side by side. 
One grid contained the locations shown on the previous four steps, while the other contained an incorrect set of locations. 
Participants were asked to press one of two buttons to choose the grid that showed the correct locations.

\subsection{Coding tasks}
\label{method::pu}
The \putasks in Experiment 1 (Python) included three conditions: problems in Python with English variables (\codee), problems in Python with Japanese variables (\codej), and problems described in English sentences (\texttt{sent}). 
The full list of problems is available on the project's code repository. 
Each participant saw 72 problems, 24 from each of the three conditions.
Any given participant saw only one version of a problem. 
Half of the problems in each condition required performing mathematical operations, and the other half required string manipulations. 
In addition, both math and string-manipulation problems varied in program structure: $\frac{1}{3}$ of the problems of each type included only sequential statements, $\frac{1}{3}$ included a \texttt{for} loop, and $\frac{1}{3}$ included an \texttt{if}/\texttt{else} statement. 
Participants were instructed to read the problem statement and press a button when they were ready to respond (the minimum reading time was restricted to 5 s and the maximum to 50 s; mean reading time was 19 s). 
Once they pressed the button, they were presented with four response options and had to indicate their response by pressing a corresponding button. 
The response screen was presented for 5 s.

Experiment 2 (\scratch) included two conditions: short programs in \scratch, and verbal descriptions of problems presented as written sentences. 
During each trial, participants were presented with a fixation cross 4 s, followed by a program for 8 s (in either code or verbal form), followed by a video that either matched or did not match that output of the program. 
Participants were instructed to indicate whether the video matched the description by pressing one of the two buttons as the video was playing. Average video length was 4.13 s (SD 1.70 s).

\subsection{Experiment procedure}
\label{method::procedure}
Each task was organized as follows - blocks contained multiple stimuli (referred to as \textit{trials} in the neuroimaging literature) from a given condition. 
Blocks from different conditions were arranged to form a \textit{run}.
The \lrspace and \mdspace localizer tasks had two blocks - a fixation block and the experiment blocks containing the localizer conditions.
Experiment blocks lasted 18 s (with 3 trials per block), and fixation blocks lasted 14 s. 
Each run (consisting of 5 fixation blocks and 16 experimental blocks) lasted 358 s. 
Each participant completed 2 runs.

For the MD localizer tasks, experimental blocks lasted 32 s (with 4 trials per block), and fixation blocks lasted 16 s. 
Each run (consisting of 4 fixation blocks and 12 experimental blocks) lasted 448 s. 
Each participant completed 2 runs.

The condition order was randomly counterbalanced across runs. 

In each run of Experiment 1 (Python), participants completed 2 trials from each of the three conditions (\codee, \codej, \sentt).
Each run included 3 fixation blocks, each of which lasted 10 s. 
One run lasted an average of 176 s (SD = 34 s). 
Each participant completed 12 runs. 

For Experiment 2 (\scratch), each run included 6 trials (three per condition), and a 10 s fixation at the beginning and end of the run.
Each run lasted an average of 184.75 s (SD 3.86 s). 
Each participant completed 4 runs. 

\subsection{Defining fROIs} 
\label{method::froi}
Function regions of interest (fROIs) were defined using the group-constrained subject-specific (GSS) approach \cite{fedorenko2010new} where a set of spatial parcels were combined with each individual subject’s localizer activation map, to constrain the definition of individual fROIs. 
The parcels mark the expected gross locations of activations for a given contrast based on prior work and are sufficiently large to encompass the extent of variability in the locations of individual activations. 

We reused parcels identified in prior works \cite{fedorenko2010new, fedorenko2013broad}. 
For the language localizer, we used a set of six parcels derived from a group-level probabilistic activation overlap map for the sentence reading (\sr) $>$ non-word reading (\nr) contrast in 220 participants. 
These parcels included two regions in the left inferior frontal gyrus (LIFG, LIFGorb), one in the left middle frontal gyrus (LMFG), two in the left temporal lobe (LAntTemp and LPostTemp), and one extending into the angular gyrus (LAngG). 

For the MD localizer, we used a set of 20 parcels (10 in each hemisphere) derived from a group-level probabilistic activation overlap map for the \texttt{hard $>$ easy} spatial task contrast in 197 participants.
The parcels included regions in frontal and parietal lobes, as well as a region in anterior cingulate.

Within each parcel, we selected the top $10$\% most responsive voxels, based on the \textsf{t}-values for the \sr $>$ \nr contrast. 
Individual-level fROIs defined in this way were then used for subsequent analyses that examined the behavior of the \lrmdspace during the \putasks.

\subsection{Data processing and analysis} 
\label{method::dataprocess}
MRI data were analyzed using SPM5. Each participant’s data were motion corrected and then normalized into a common brain space (the Montreal Neurological Institute (MNI) template) and resampled into 2mm isotropic voxels. 
The data were then smoothed with a 4mm FWHM Gaussian filter and high-pass filtered (at 200s). 
Effects were estimated using a General Linear Model (GLM) in which each experimental condition was modeled with a boxcar function (modeling entire blocks) convolved with the canonical hemodynamic response function (HRF).
%\shash{some more details and references perhaps.}

\section{Additional Results}
We provide additional results from our data analysis of the \lrmdspace.
\subsection{fROI responses - \mdspace}
\label{froi_results}
\begin{figure}[ht]
	\begin{center}
		\includegraphics[width=0.9\linewidth ]{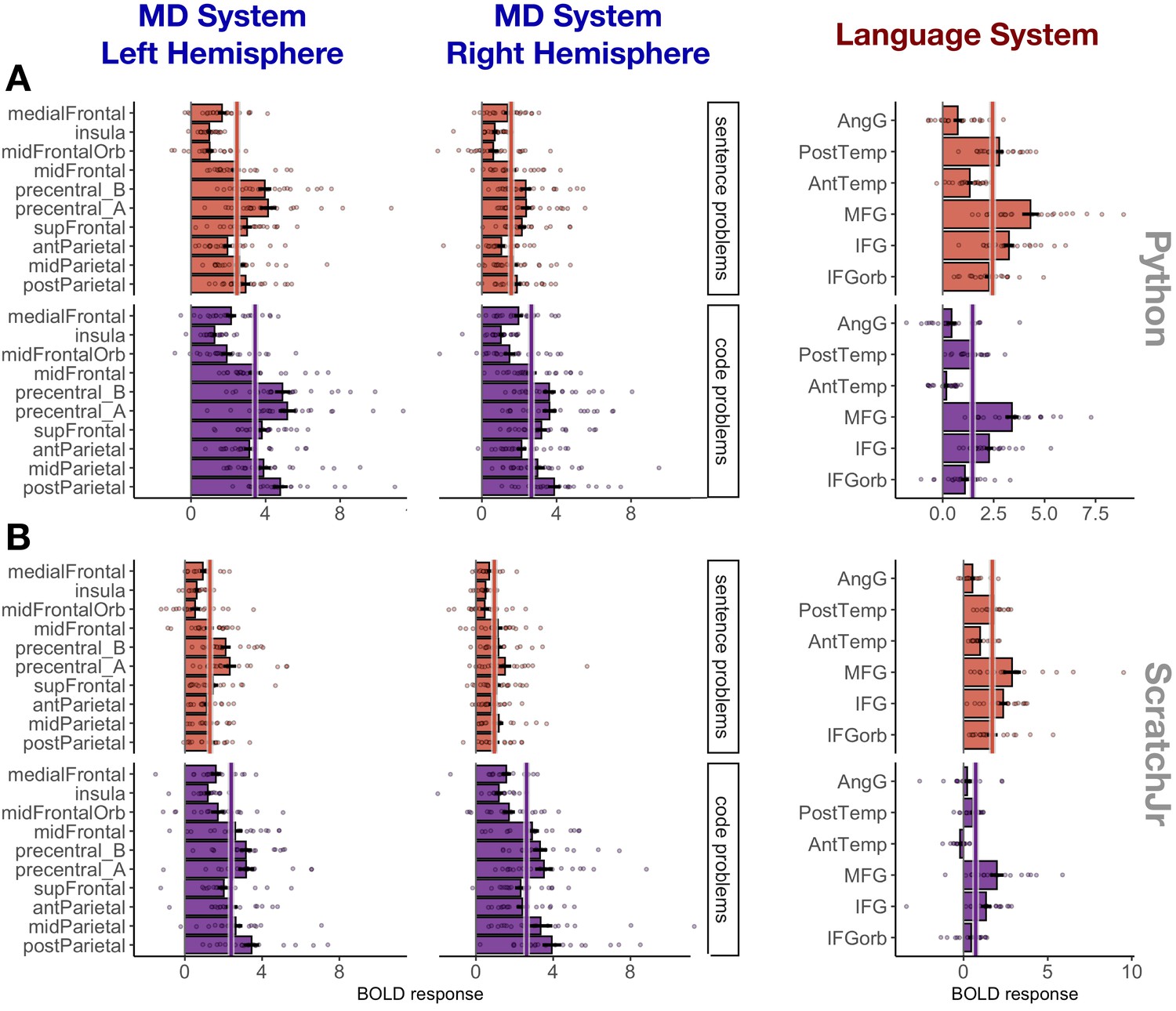}
	\end{center}
	\caption{Activations of fROIs to \texttt{code} and \texttt{sent} conditions in the \mdspace - left hemisphere, \mdspace - right hemisphere, and the \lrspace.}
	\label{results2}
\end{figure}

\begin{figure}[ht]
	\begin{center}
		\includegraphics[width=0.9\linewidth ]{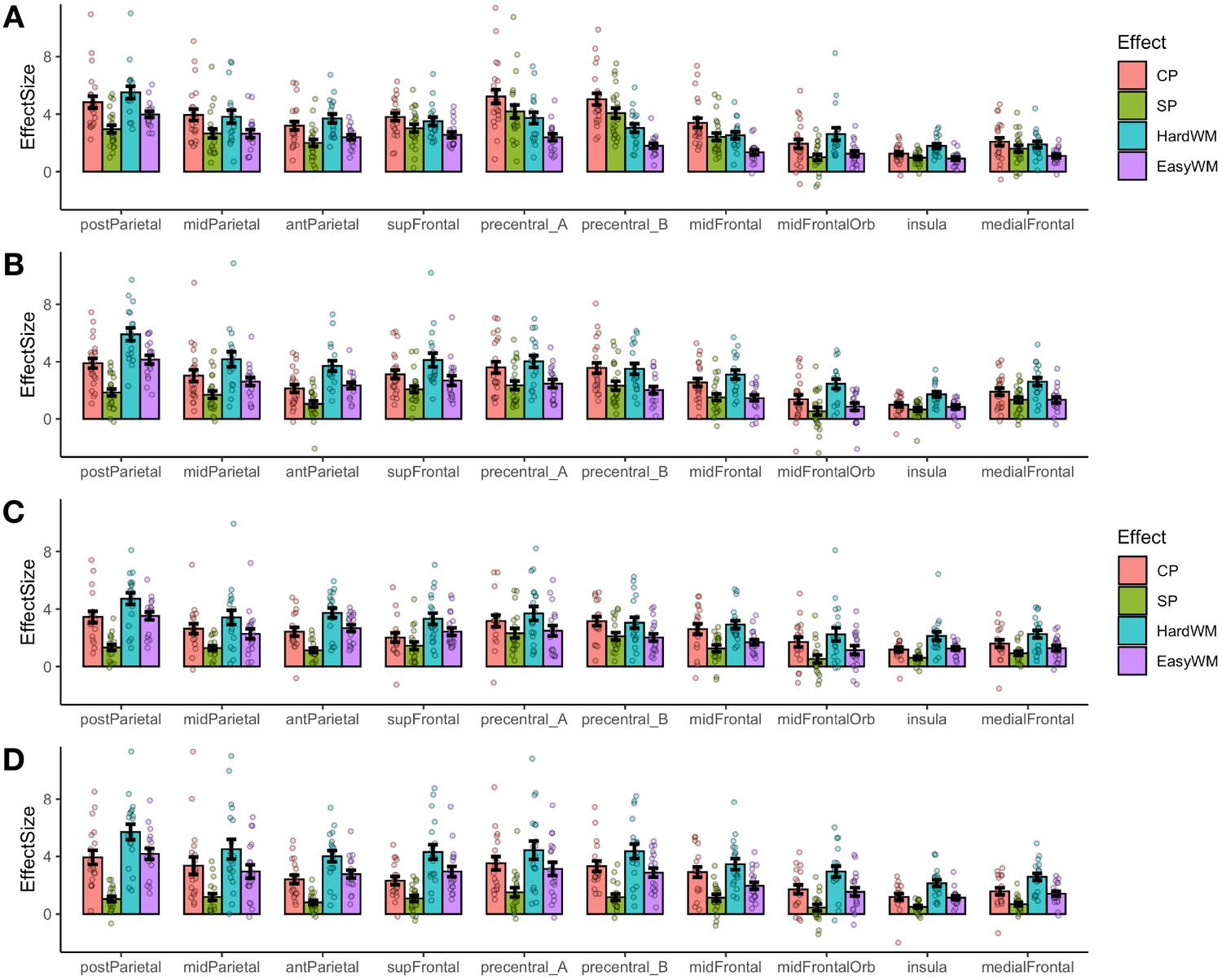}
	\end{center}
	\caption{Activations of the \mdspace fROIs to \texttt{code} and \texttt{sent} conditions, contrasted against the \texttt{easy} and \texttt{hard} \mdspace localizer tasks. (A) Experiment 1, Python; left hemisphere fROIs; (B) Experiment 1, Python; right hemisphere fROIs; (C) Experiment 2, ScratchJr; left hemisphere fROIs; (D) Experiment 2, ScratchJr; right hemisphere fROIs.}
	\label{results3}
\end{figure}
An analysis of activity within individual regions within the \mdspace showed that $17$ of the $20$ fROIs in the Python experiment, and  $14$ of the $20$ fROIs in the \scratch experiment responded significantly more strongly to code problems than to sentence problems (Figure \ref{results2}).
This demonstrates code processing is broadly distributed across the \mdspace and is not localized to a particular subset of regions within it.

We evaluate for \textit{selectivity} to \codecomp by measuring responses of code problems to hard working memory localizer task for the \mdspace. 
Figure \ref{results3} plots activations in the various regions of the \mdspace.
We find none for \scratch, and three regions in the frontal lobe (precentral-A, precentral-B, midFrontal) which exhibit stronger responses to code problems.
However, the magnitude of \texttt{code} $>$ \texttt{sent} in these regions (\delbeta$=1.03, 0.95, 0.97$) was comparable to the mean magnitude across all \mdspace fROIs (average \delbeta$=1.03$), suggesting that the high response was caused by the underlying \codecontent rather than \codecomp.

\subsection{Behavorial Results}
\label{beh_results}
\begin{figure}[hbpt]
	\begin{center}
		\includegraphics[width=0.9\linewidth ]{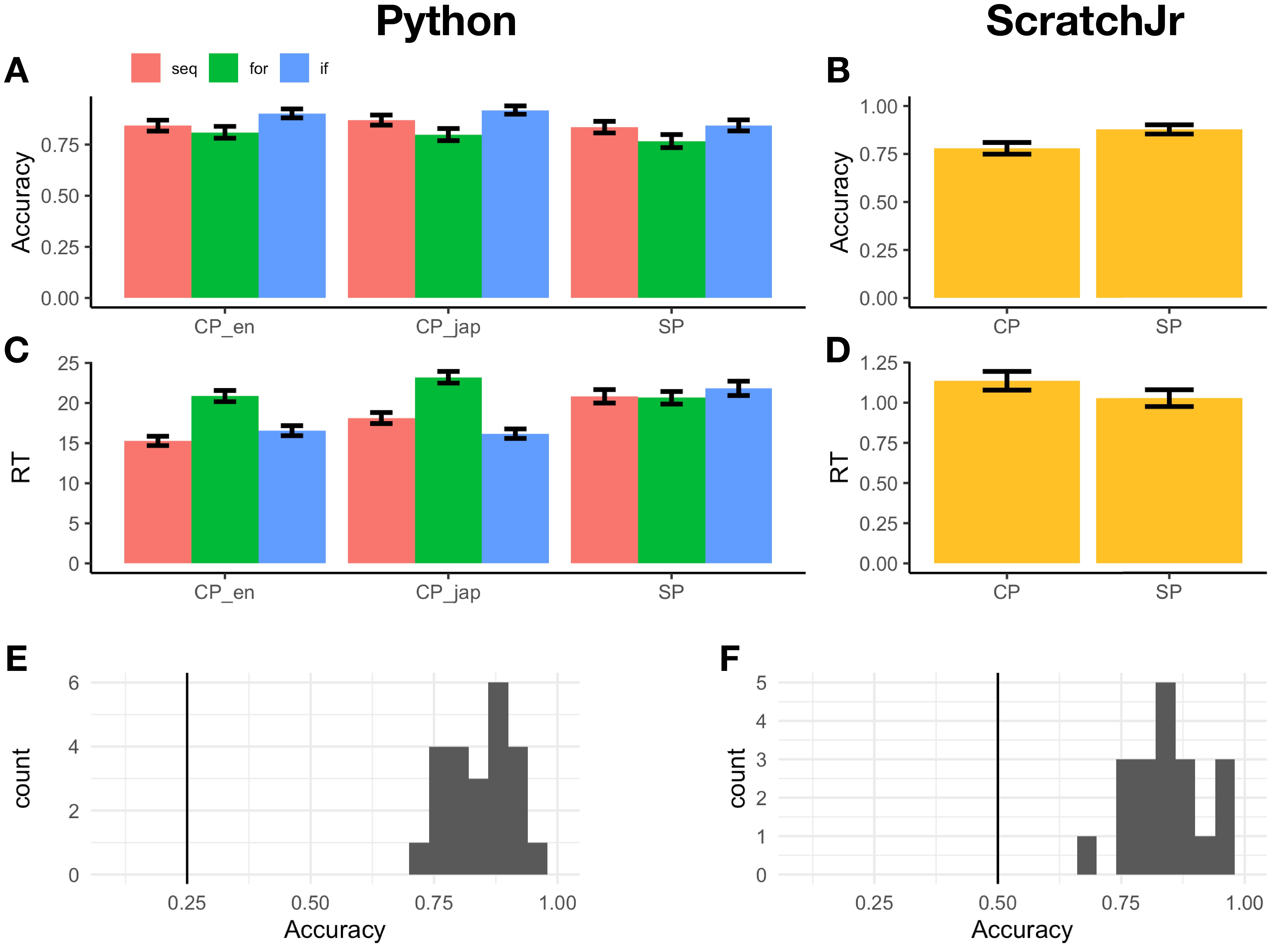}
	\end{center}
	\caption{Response accuracies and reaction times to code-related stimuli}
	\label{results4}
\end{figure}
Figure \ref{results4} presents response accuracies and reaction times to code-related stimuli presented to our participants in the imaging scanner.
Participants in Experiment 1 (Python) had a 99.6\% response rate, with an 85\% accuracy on average on code problems (Figure \ref{results4}.A). 
Figure \ref{results4}.E shows the histogram of response accuracies of participants in Python.

Participants in Experiment 2 (ScratchJr) had a 98.6\% response rate, with 79\% accuracy on average on code problems (Figure \ref{results4}.B).
These results demonstrate that participants were proficient in the relevant programming language and engaged with the task.
Figure \ref{results4}.F shows the histogram of response accuracies of participants in \scratch.